\documentclass{epl}
\usepackage{graphicx}
\usepackage{amssymb}
\usepackage{amsmath}

\begin{document}

\title{Entanglement transition of elastic lines in
       a strongly disordered environment} 
\author{Viljo Pet\"aj\"a\inst{1}
\and Mikko Alava\inst{1,2}
\and Heiko Rieger\inst{3}}


\institute{
\inst{1} Helsinki University of Techn., Lab. of Physics, 
P.O.Box 1100, 02015 HUT, Finland\\
\inst{2} SMC-INFM, Dipartimento di Fisica,
Universit\`a ``La Sapienza'', P.le A. Moro 2
00185 Roma, Italy\\
\inst{3} Theoretische Physik, Universit\"at des Saarlandes, 
66041 Saarbr\"ucken, Germany}

\maketitle
\newcommand{\bc}{\begin{center}}
\newcommand{\ec}{\end{center}}
\newcommand{\be}{\begin{equation}}
\newcommand{\ee}{\end{equation}}
\newcommand{\ba}{\begin{array}}
\newcommand{\ea}{\end{array}}
\newcommand{\beqn}{\begin{eqnarray}}
\newcommand{\eeqn}{\end{eqnarray}}

\begin{abstract}
We investigate by exact optimization the geometrical
properties of three-dimensional elastic line systems with point
disorder and hard-core repulsion.  The line 'forests' become
{\it entangled} due to increasing line wandering
as the system height is increased, at fixed line density.
There is a transition height at which a cluster of pairwise entangled
lines spans the system, transverse to average line orientation.
Numerical evidence implies that the phenomenon is in the ordinary percolation
universality class. Similar results are obtained for an ensemble
of flux lines obeying random walk -dynamics.
\end{abstract}

The topological entanglement of line-like elastic objects is an
important physical phenomenon for a number of interacting many-body
systems. Examples are the entanglement of magnetic flux lines in
high-T$_c$ superconductors in the mixed phase \cite{nelson,blatter} or
of polymers in materials like rubber
\cite{polymer-review}. The degree of entanglement usually
manifests itself in various measurable properties like stiffness or
shear modulus in the case of polymers and in transport or dynamical
properties for magnetic flux lines in superconductors. 

In the latter case the entanglement is caused either by thermal
fluctuations \cite{nelson,thermal} or sufficiently strong quenched
disorder through lattice defects or pinning centers
\cite{obukhov,hwa,lopez} and varies with the line density, i.e.\
the strength of the applied magnetic field.
A strongly disordered high-$T_c$ superconductor is in
a vortex glass state \cite{vortexglass}, which is superconducting (or
has at least an extremely small resistivity) and in which the
Abrikosov flux line lattice is destroyed as in the non-superconducting
vortex liquid state. At weaker disorder or
at smaller magnetic fields a transition to a Bragg glass phase
\cite{bragg-glass}, in which the positions of the flux lines show
quasi-long-range order, takes place. This has been observed
experimentally \cite{exp} and in simulations \cite{sim}. In contrast
to this (quasi)-ordered solid, the lines in the vortex glass phase are
expected to be strongly entangled \cite{samokhin}.  Since the vortex
glass proposed in \cite{vortexglass} corresponds to glassy order in
the phase of the superconducting order parameter it is not a priori
clear that an entangled phase corresponds to this, it
could be another type of glassy phase, or smoothly connected to a
liquid. In any case an entangled phase will be different from the
Bragg glass phase, and an elastic description, the usual theoretical
starting point \cite{bragg-glass}, is inappropriate
due to the drastic increase in the transverse line fluctuations.
An understanding of such topologically highly
non-trivial states including the conditions under which it occurs
is highly desirable.

Here we study a model for an ensemble of elastic lines
exposed to strong point disorder and focus on the transition
from independent lines to entangled bundles of lines.
Due to the disorder potential the lines make
transverse excursions even at zero temperature, and
wind around each other. We introduce a computationally convenient
measure for the entanglement of two lines via their winding angle and
study clusters of mutually entangled lines. The physical
properties of the line system are expected to change drastically (if
compared with an unentangled, Bragg-glass like case) with the size of
these clusters. Beyond a particular height of the
system, dependent on the line density $\rho$, a percolating bundle
of entangled lines is formed. It spans the transverse system
size --- representing a disorder induced braided flux phase. Thus, in
accordance with an observation by Nelson for the thermal case
\cite{nelson}, in sufficiently thin samples the lines
behave like disentangled rods (still pinned by the disorder),
whereas for thick enough systems there is an
entanglement transition. This takes place also if the line density 
is varied - in high-$T_c$ superconductors - by 
the magnetic field, though our model is strictly appropriate at
relatively low densities when the line-line interactions are weak.
This transition may be visible in experiments 
\cite{exp} with a Bragg glass phase and a vortex glass
phase, in which flux lines are expected to be entangled (although
experimentally line entanglement might be difficult to measure). The
results are contrasted with a simulation of a (2+1)-dimensional set of
random walkers - mimicking thermally fluctuating flux lines with 
hard core repulsion - with a similar entanglement transition.

The three-dimensional  disordered environment has 
lines living on the
bonds of a simple cubic lattice with a lateral width $L$ and a longitudinal
height $h$ ($L\times L\times h$ sites) with free
boundary conditions in all directions. Each line starts and ends
at an arbitrary position on the bottom respective top planes.
The number $N$ of lines is fixed
by a prescribed density $\rho=N/L^2$. The lines have
hard-core interactions: the
configuration is specified by bond-variables $n_i=1$,
indicating a line segment occupying bond $i$, and $n_i=0$
for vacant bonds. Note that we allow line
crossing at the lattice sites, and thus the line
identification based upon a bond configuration is not unique.

We model the disordered environment by assigning a random (potential)
energy $e_i\in[0,1]$ (uniformly distributed) to each bond $i$.
The total energy of a configuration is given by
\be
H=\sum_i e_i n_i\;.
\label{ham}
\ee
An elastic energy term is intrinsically present in this model (i.e.\
is generated upon coarse graining \cite{opt-review})
since the positivity of
the energies $e_i$ disfavor in general large transverse excursions of
the lines. An equivalent continuum version is 
\be
{\mathcal H} = {\displaystyle\sum_{i=1}^N} 
{\displaystyle\int_0^h dz}
\Bigl\{ \frac{\gamma}{2}\left[\frac{d{\bf r}_i}{dz}\right]^2
+\sum_{j(\ne i)}V_{\rm int}[{\bf r}_i(z)-{\bf r}_j(z)]
+V_r[{\bf r}_i(z),z]\Bigr\}\;.
\label{cont}
\ee
where ${\bf r}_i(z)$ denotes the transversal coordinate at
longitudinal height $z$ of the $i$-th flux line and 
where, in order to version (\ref{ham})
the interactions $V_{\rm int}[{\bf r}_i(z)-{\bf r}_j(z)]$ have to be
short ranged (i.e.\ in case of flux lines the screening length small
compared to the average line distance) or just hard core repulsive,
and the random, $\delta$-correlated disorder potential $V_r[{\bf r}_i(z),z]$
has to be strong compared to the elastic energy ($\propto \gamma$).

At low temperatures the line configurations will be dominated by the
disorder and thermal fluctuations are negligible. We
focus ourselves to zero temperature, the ground state
of the energy (\ref{ham}). The computation thereof
is a non-trivial task and is done by applying the polynomial 
Dijkstra's shortest path algorithm, successively
on a residual graph \cite{flux,opt-review}.

\begin{figure}[t]
\begin{center}
    \begin{minipage}[t!]{0.58\columnwidth}
      \onefigure[width=0.7\columnwidth]{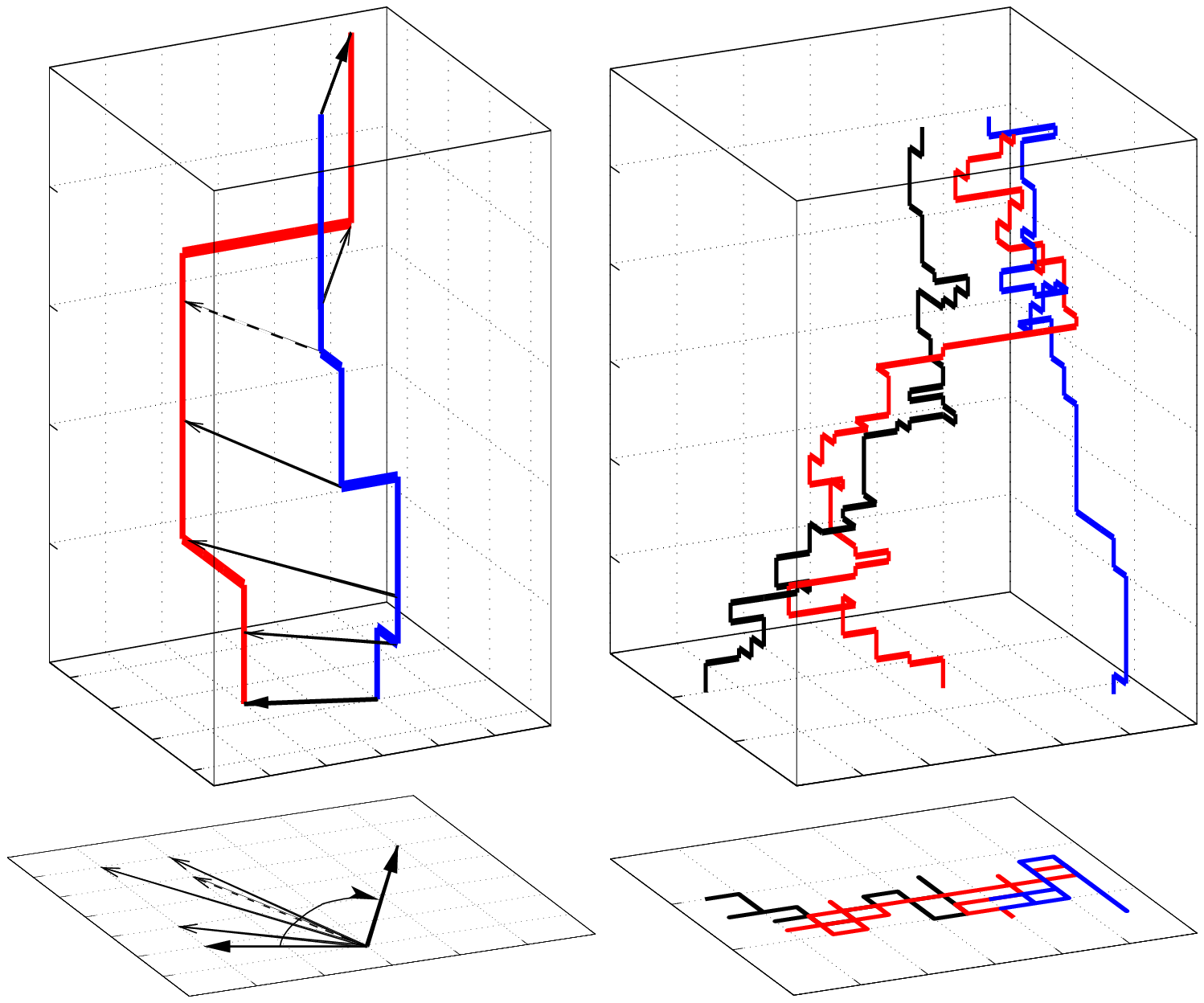}
    \end{minipage}\hfill
    \begin{minipage}[t!]{0.4\columnwidth}
      \caption{\label{fig1} Left: Definition of the winding angle of two
	flux lines: For each z-coordinate the vector connecting the two lines
	is projected onto that basal plane. $z=0$ gives
	the reference line with respect to which the consecutive vectors for
	increasing $z$-coordinate have an angle $\phi(z)$. Once $\phi(z)>2\pi$
	the two lines are said to be entangled. Right, top: A configuration of
	three lines that are entangled.  Right, bottom: The projection of the
	line configuration on the basal plane, defining a connected cluster.}
    \end{minipage}
  \end{center}
\onefigure[width=0.66\columnwidth]{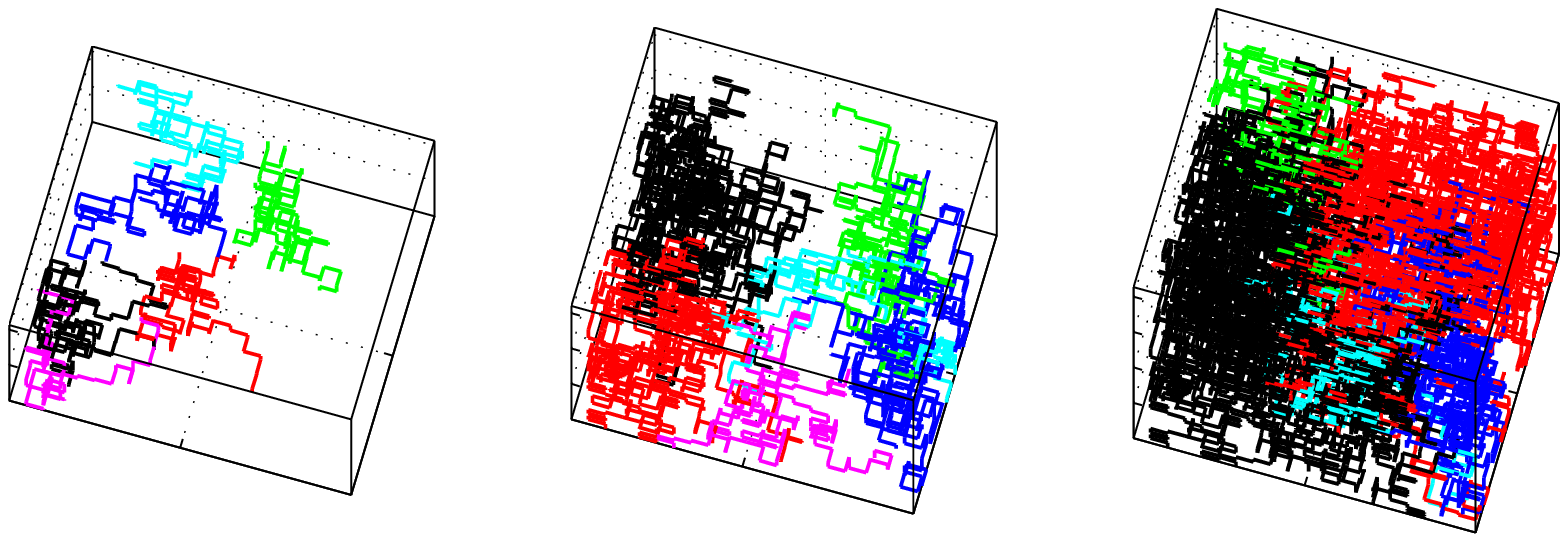}
\vspace{-0.4cm}
\caption{\label{fig2} Line configurations for different heights $h$
(from left to right: $h=64$, $96$, $128$), the lateral size 
$L=20$, the line density is $\rho=0.3$. Only the largest line bundles
are shown, indicated by a varying grey scale. Black denotes the
largest cluster, which eventually percolates.}
\end{figure}

The line configuration is then analyzed by computing the
winding angle of all line pairs as indicated in Fig.\ \ref{fig1} (c.f.\
\cite{drossel}). We define two lines to be {\it entangled} when this
winding angle becomes larger than $2\pi$. This
measures entanglement from the topological perspective
\cite{samokhin}, arising from
the requirement that an entangled pair of lines can not be separated
by a linear transformation in the basal plane (i.e. the lines
almost always would cut each other, if one were shifted). The precise 
definition of entanglement is not of major relevance, this being
the computationally easiest.

Sets or {\it bundles} of pairwise entangled lines are formed so 
that a line belongs to a bundle if it is entangled at least with one other
line in the set. Such bundles are
spaghetti-like --- i.e. topologically complicated and knotted sets
of one-dimensional objects. Their size distribution is found 
by a projection on the basal plane, (Fig.\ \ref{fig1}), 
a bundle projecting onto a connected cluster.
The probability for two-line entanglement increases
with system height. Consequently the
bundles and their projections (clusters) grow with $h$.
as exemplified in Fig.\ \ref{fig2}. For
the largest height the largest cluster spans from one side
of the system to the other, i.e.\ it {\it percolates}.
Hence, for a given line density $\rho$ we expect that for system
heights larger than a finite critical value $h_c$ a spanning 
entangled bundle occurs, containing an infinite number of lines in
the limit $L\to\infty$. We denote this an {\it entanglement transition}.
In the projection plane this is like a percolation transition,
whose properties are investigated next.

\begin{figure}[t]
\onefigure[width=0.47\columnwidth,angle=270]{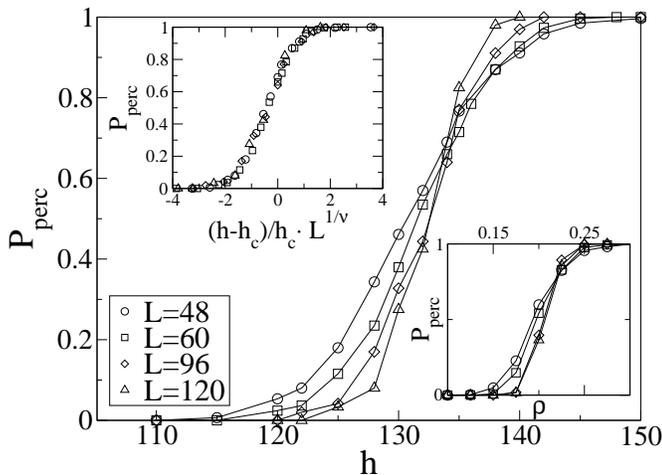}
\vspace{-0.55cm}
\caption{\label{fig3} Percolation probability for different lateral
system sizes $L$ as a function of the system height $h$, the line
density is $\rho=0.3$. {\bf Upper inset:} Scaling plot of the data 
with $h_c=134$ and $\nu=4/3$.
{\bf Lower inset:} Behavior for $h=150$, with $\rho$ as the control
parameter.}
\end{figure}

The numerical data has been obtained by averaging over 
up to $10^3$ realizations of the random potentials $e_i$ in
(\ref{ham}) and the resulting statistical error
is in all cases smaller than the symbol size.
We studied different line densities
between $\rho=0.1$ and $\rho=0.5$, but present for brevity only 
data for $\rho=0.3$. The $\rho$-dependence of the
non-universal quantities like $h_c$ is discussed later --- 
the exponents we report are universal.

\begin{figure}
\onefigure[width=0.47\columnwidth,angle=270]{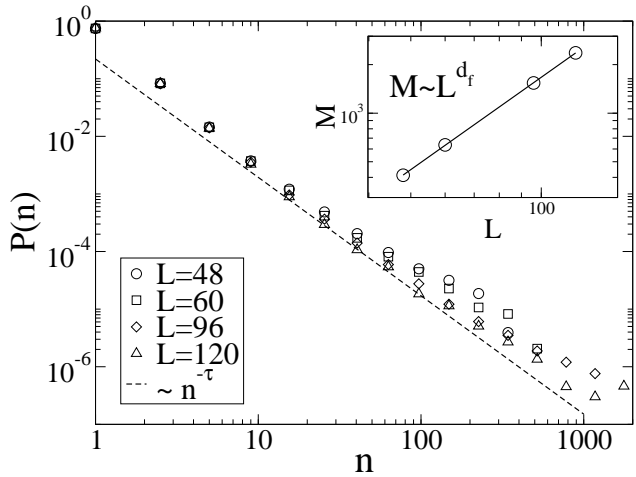}
\vspace{-0.55cm}
\caption{\label{fig4} Cluster size distributions for different lateral
system size at the percolation height $h_c=134$ in log-log plot. One
sees the finite size effects in the tails, the broken line has slope
$\tau=2.06$ and fits the asymptotic ($L\to\infty$) $P(n)\propto
n^{-\tau}$. {\bf Inset:} Log-log-plot of the mass (number of entangled
lines) of the percolating cluster at $h_c=134$
as a function of the lateral size $L$. The straight line has
slope $d_f=1.90$, the fractal dimension of percolation.}
\end{figure}

In Fig.\ \ref{fig3} we show the  probability $P_{\rm perc}$
of the clusters of entangled bundles, to percolate 
as a function of the height $h$ of the system. The
curves for different lateral system sizes $L$ intersect at $h_c$,
which gives our estimate for $h_c(\rho=0.3)=134$. The upper inset shows a
scaling plot according to 
\be
P_{\rm perc}=p(L^{1/\nu}\delta)
\ee
with $\delta=(h-h_c)/h_c$ the reduced distance from the critical height
and $\nu=4/3$ the correlation length exponent. The finite size
corrections for smaller system sizes than those shown are
larger due to single lines percolating.
The lower inset demonstrates that the density $\rho$ can also
be used as a control parameter.
$\nu=4/3$ is the exponent of conventional bond percolation
in two dimensions. Thus we conclude that the transition
is in the universality class of conventional 2$d$ percolation. Other
quantities confirm this result: Fig.\ \ref{fig4} shows the cluster
size distribution $P(n)$ at $h=h_c$ for various 
$L$ and we find it that it approaches 
\be
P(n)\propto n^{-\tau}\quad{\rm with}\quad\tau=187/91\approx2.055
\ee
in the limit $L\to\infty$. The inset shows the mass (i.e.\ number of
entangled lines) of the percolating bundle at $h_c$, which fits well
to 
\be
M\propto L^{d_f}\quad{\rm with}\quad d_f=91/48\approx1.896
\ee
Both exponents,
the cluster distribution exponent $\tau$ and the fractal dimension
$d_f$ are identical to conventional bond percolation and
the order parameter exponent, $\beta=5/36$ also fits the data reasonably well.

\begin{figure}
\onefigure[width=0.47\columnwidth,angle=270]{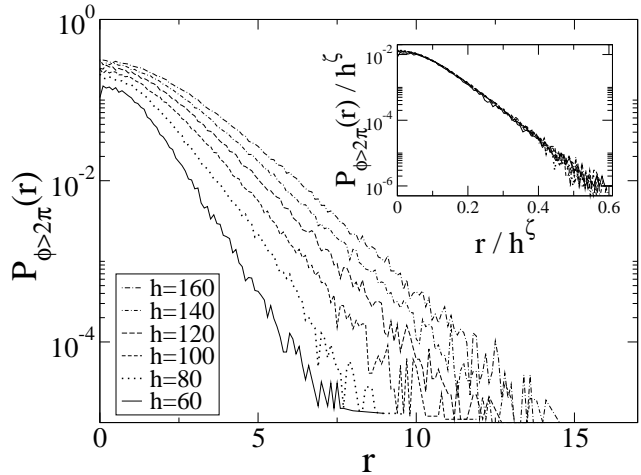}
\vspace{-0.55cm}
\caption{\label{fig5} Probability $P_{\phi>2\pi}(r)$ that two lines at
a mass center distance $r$ have a winding angle larger than $2\pi$,
i.e. are entangled. Inset: Scaling plot,
 according to $P_{\phi>2\pi}(r)=h^\zeta\tilde{p}(r/h^\zeta)$.
$\zeta=0.62$ is the single line roughness exponent. The lateral system
size is $L=60$, since $r\ll L$ there is no finite $L$ effect.}
\end{figure}

This result is somewhat surprising, since one could expect
the entanglement of two lines to have
correlations. These, if long ranged,
change the percolation universality class
\cite{weinrib}. Why this is not true is seen
from the probability for two lines with mass center distance $r$
to be entangled, $P_{\phi>2\pi}(r)$ (Fig.\
\ref{fig5}). It decays exponentially with $r$, implying
for two-line-entanglement only short range correlations. It
scales with the transverse fluctuations of a
single line $h^\zeta$, where $\zeta=0.62$ is the single-line
roughness exponent in 3$d$,
\cite{1line3d,1line-review}, so that
$P_{\phi>2\pi}(r)=h^\zeta\tilde{p}(r/h^\zeta)$.
It is noteworthy that the ``really entangled''
lines follow a scaling stemming from single-line behavior.
Three-line correlations (e.g.\ the probability of a
third line to be entangled with one or two others, given that these
two are entangled or not), are also only short-ranged.

The critical height, $h_c$, above which an extensive fraction of lines
is entangled 
varies non-monotonically with the line density $\rho$. We expect
that $h_c\to\infty$ in both limits $\rho\to0$, i.e.\ very dilute
system, and $\rho\to1$, very dense systems. In the former case
because the line fluctuations $w\propto h^\zeta$ 
should be of the same order of magnitude as the typical 
line-to-line distance $d=1/\sqrt{\rho}$. For this 
the height has, obviously, to be large enough (see
Fig.\ \ref{fig3} in which $h=150$ was chosen to study 
percolation with $\rho$).
In the opposite limit of a dense system the effective stiffness starts to 
increase due to the hard core interactions (ie. a lack of
vacancies, or unoccupied bonds).
Hence $h_c(\rho)$ has a minimum between these two
limits, in practice around $\rho=0.4$.
Our estimates for the critical exponents are independent of
$\rho$, but the quality of the data detoriates due to fewer samples.
We expect that for all $\rho$ the entanglement transition is in the 
percolation universality class.

\begin{figure}
\onefigure[width=0.47\columnwidth,angle=270]{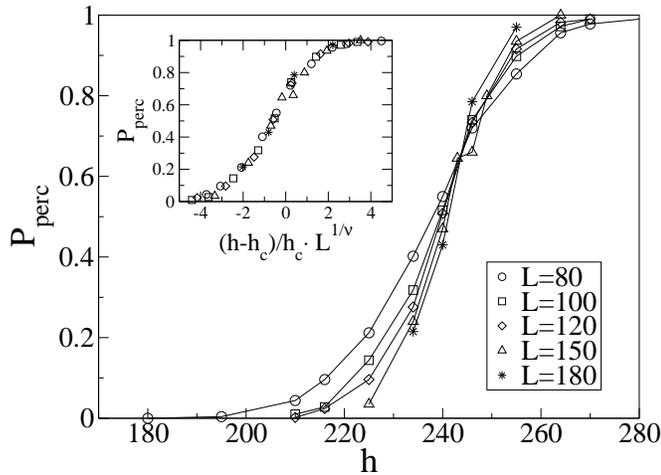}
\vspace{-0.55cm}
\caption{\label{rws} Percolation probability for the random walk
--case, as a function of the system height $h$, the line
density is $\rho=0.1$. {\bf Inset:} Scaling plot of the data 
with $h_c=244$ and $\nu=4/3$.}
\end{figure}

The previous picture of the development of entanglement underscores
the weak correlations in the line ensemble - the character of
the ground state enters mostly via single-line roughness exponents.
An analogous question can be put forth in the hypothetical case of
$N = \rho L^2$ flux lines, that roughen thermally but do not interact
except via hard-core expulsion.
In this thermal case flux lines are modeled by an configuration of 
simultaneously evolved non-overlapping random walks on the tilted cubic 
lattice where $z$-direction represents the time.
At each step $z\to z+1$ all walks in a randomly chosen order make a 
step using non-occupied bonds.
Results are shown in Fig. \ref{rws}, and can be compared with
Fig. \ref{fig3}. The data collapses as before, and does so for 
$P_{\phi>2\pi}(r)$ with the single-random walk roughness exponent 
(1/2). 

To conclude we studied the entanglement transition of hard core
repulsive lines in a disordered environment, driven by the
system thickness or the line density. We find two transitions or
a non-monotonic behavior of $h_c(\rho)$, for sufficiently thick samples. 
The one at low densities may be, in the presence of strong point
disorder, located close to the the Bragg to vortex glass transition
\cite{exp,sim}. A signature of this would be a
dependence of the order-disorder phase boundary on the sample thickness.
Above the transition an extensive line bundle is formed. 
It presents a topological obstacle
for also those lines that are not part of it, due to the energy
cost of vortex-vortex cutting. In the proximity of the transition
there is a wide variety of bundle sizes, in full analogy with
usual percolation. This has implications for {\em bundle depinning},
a collective effect that depends on the bundle size \cite{yeh},
if an external current is applied, and thus on vortex flow both
for bundles and single vortices.
In analogy with thermal entanglement \cite{nelson} one expects
also the critical current to increase \cite{beek}.

It would be interesting to study the modifications of
the entanglement transition due to splayed line defects \cite{hwa}, or
other forms of correlated disorder. Moreover, it would be highly
desirable to include long-range interactions among the lines, since
only in their presence a Bragg glass phase on the low disorder or low
line density side will occur. The addition of thermal fluctuations,
neglected here, will lead to another interesting transition,
from an entangled vortex glass to an entangled vortex liquid.

\acknowledgments
The support of the European Science
foundation (ESF) SPHINX network, 
the Deutsche Forschungsgemeinschaft (DFG), and (MJA, VIP)
the Center of Excellence program of the Academy of Finland
is acknowledged.

\end{document}